\documentclass[prc,showpacs,amsmath,showkeys,twocolumn,floatfix]{revtex4} 

\usepackage{amsmath}
\usepackage{graphicx}
\usepackage{color}

\def\gtorder{\mathrel{\raise.3ex\hbox{$>$}\mkern-14mu
	\lower0.6ex\hbox{$\sim$}}}
\def\ltorder{\mathrel{\raise.3ex\hbox{$<$}\mkern-14mu
	\lower0.6ex\hbox{$\sim$}}}

\def\lsim{\mathrel{\rlap{\lower4pt\hbox{\hskip1pt$\sim$}}
    \raise1pt\hbox{$<$}}}
\def\gsim{\mathrel{\rlap{\lower4pt\hbox{\hskip1pt$\sim$}}
    \raise1pt\hbox{$>$}}}

\def \EG{{\it e.g.,}}
\def \IE{{\it i.e.,}}
\def \GDR{{\scriptscriptstyle{GDR}}}
\def \CQM{{\scriptscriptstyle{CQM}}}
\def \beqn{\begin{eqnarray}}
\def \eeqn{\end{eqnarray}}

\def \bea{\begin{eqnarray}}
\def \beq{\begin{equation}}
\def \eea{\end{eqnarray}}
\def \eeq{\end{equation}}
\def \nn{\nonumber}
\def \bwt{\begin{widetext}}
\def \ewt{\end{widetext}}

\begin{document}
\title{Compton Scattering and Photo-absorption Sum Rules on Nuclei}



\author{Mikhail Gorchtein}
\affiliation{Department of Physics and Center for Exploration of Energy
and Matter, Indiana University, Bloomington, IN 47405, USA}
\author{Timothy  Hobbs}
\affiliation{Department of Physics and Center for Exploration of Energy
and Matter, Indiana University, Bloomington, IN 47405, USA}
\author{J. Timothy Londergan} 
\affiliation{Department of Physics and Center for Exploration of Energy
and Matter, Indiana University, Bloomington, IN 47405, USA}
\author{Adam P. Szczepaniak}
\affiliation{Department of Physics and Center for Exploration of Energy
and Matter, Indiana University, Bloomington, IN 47405, USA}

\begin{abstract}
We revisit the photo-absorption sum rule for real Compton
scattering from the proton and from nuclear targets. In analogy with the 
Thomas-Reiche-Kuhn sum rule appropriate at low energies, we propose a new 
"constituent quark model" sum rule that relates the integrated strength of hadronic
resonances to the scattering amplitude on constituent quarks. We study the 
constituent quark model sum rule for several nuclear targets. In addition we 
extract the $\alpha=0$ pole contribution for both proton and 
nuclei. Using the modern high energy proton 
data we find that the $\alpha=0$ pole contribution differs
significantly from the Thomson term, in contrast with the 
original findings by Damashek and Gilman. 
\end{abstract}
\date{\today}

\maketitle


\section{INTRODUCTION}
\label{sec:intro}

Compton scattering on composite objects has served as a valuable
tool for studying internal structure of nuclei and nucleon. At very low photon energy, 
electromagnetic waves are scattered without absorption and solely probe the
macroscopic properties of the target as a whole, \EG~its mass and 
electric charge, and the scattering amplitude is determined by the 
classical Thomson limit. As the photon energy $\nu$ is 
increased above the absorption threshold, the internal structure of the 
target is revealed. The atomic, nuclear and hadronic physics domains roughly 
correspond to keV, MeV and GeV photon energies, respectively, and 
the three orders of magnitude difference between neighboring 
domains indicates that the dynamics of nuclei are well separated from 
those of quarks so that each can be clearly identified.

In this work, we compare photon scattering from nuclei with photon scattering 
off the individual nucleons.  At energies beyond the nuclear absorption range, 
\IE~of the order of tens of MeV's, the interaction time between the photon and 
the target is much shorter then that between individual nucleons, and in 
this regime the scattering amplitude is determined by Thomson scattering on 
independent nucleons.  By applying the optical theorem this relation can be 
made quantitative, and this leads to a sum rule relating the low (MeV range) 
and medium energy (tens of MeV's) scattering amplitudes on a nucleus to the 
total photo-absorption cross section~\cite{Levinger:1960} in this energy 
range. Following the success of the constituent quark model of low energy 
hadron structure one would expect that this nuclear sum rule might be extended 
to cover the energy range between pion production threshold (roughly 100 MeV) 
to above the range of nucleon resonances (a few GeV). Such an extended 
`constituent quark model' sum rule would therefore relate the photo-absorption 
cross section on a nuclear target to the difference between the low-energy 
scattering amplitude on the nucleus and the scattering amplitude describing 
photon interactions with individual constituent quarks. In this paper we 
will derive a finite energy sum rule involving constituent quarks. We will 
investigate its validity by including nuclear photo-absorption data above the 
pion production threshold which is now available for a wide range of nuclear 
targets.   
   
As the energy range is further increased, we expect that the QCD structure of 
the constituent quarks will be resolved, eventually revealing scattering on 
point-like quarks. 
\begin{figure}
\includegraphics[height=2cm]{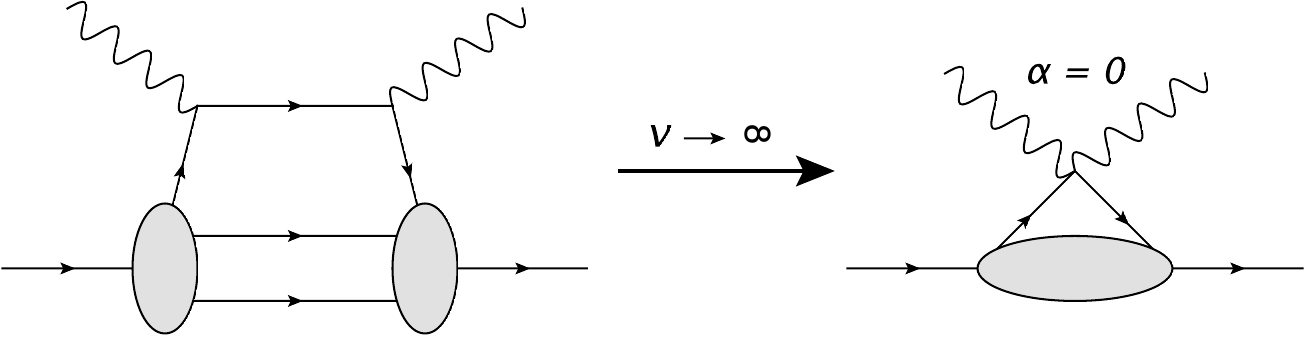}
\caption{The fixed-pole contribution to the Compton amplitude may
  arise due to an effective local two-photon coupling to elementary
  constituents within the proton.}
\label{fig:diag}
\end{figure}
Thus at asymptotically high energies point-like interactions  
involving photons can contribute an energy-independent constant to the 
amplitude, which corresponds to a Regge pole at $\alpha=0$ 
~\cite{Damashek:1969xj,j0r,j01,Creutz:1968ds,Close:1971ed,Zee:1972nq, Brodsky:1971zh, Brodsky:1973hm,Brodsky:2008qu}, 
as is shown schematically in Fig.~\ref{fig:diag}. In the presence of other 
poles in the right-half of the angular momentum plane, the $\alpha=0$
pole \cite{footnote}
produces a sub-leading contribution to the scattering amplitude.  Nevertheless 
since contributions from leading poles with $\mbox{Re}\alpha>0$ can be 
determined by fitting a Regge-type amplitude to the high energy data, it might 
be possible in principle to extract the residual $\alpha=0$
pole. In the past 
this procedure has been carried out for the proton 
\cite{Damashek:1969xj,Dominguez:1970wu,Shibasaki:1971,Kugler:1971,Tait:1972}
and the deuteron \cite{Dominguez:1972}. In this work we 
re-examine the procedure for extracting the $\alpha=0$ pole by including in 
our analysis data at very high energies that were not available when the 
original analysis was performed in 1969. We will show that, with this new 
data, one can unambiguously extract the $\alpha=0$ pole. We also examine 
possible $\alpha=0$ pole contributions to Compton scattering on heavier 
nuclei. 

Our paper is organized as follows. In the next section we focus on the energy 
range up to the pion production threshold and we summarize the derivation 
of the nuclear photo-absorption sum rule, also referred  to as the 
Thomas-Reiche-Kuhn (TRK) 
sum-rule~\cite{Thomas:1925a,Thomas:1925b,Kuhn:1925,Levinger:1960}.  
In Sec. ~\ref{sec:j_form} we generalize the TRK sum rule to cover energies 
beyond the pion threshold where we include hadronic resonances and we test the 
validity of a new finite-energy sum rule based on a constituent quark 
picture. Finally we consider energies above the GeV range. We discuss the 
implications of scattering on QCD partons and we extract the $\alpha=0$ pole 
contribution to scattering at asymptotic energies for various nuclear 
targets. Our summary and conclusions are presented in Sec.~\ref{summary}.  

\section{Nuclear photo-absorption at low energies} 

The spin-averaged forward Compton scattering amplitude $T(\nu)$ satisfies a 
once-subtracted dispersion relation where the subtraction constant at $\nu=0$  
is determined by the classical Thomson limit,  

\begin{eqnarray} 
\label{eq:DR}
& & {\rm Re}T(\nu)=  -\frac{Z^2}{A^2} \frac{\alpha}{M_N}
+\frac{\nu^2}{\pi} \int_{0}^\infty\frac{d\nu'^2
}{\nu'^2(\nu'^2-\nu^2)} {\rm Im}T(\nu') \nonumber \\
\end{eqnarray} 
where the integral in Eq.~(\ref{eq:DR}) is understood in terms of its 
principal value. To facilitate easier comparison between different nuclei we 
have normalized $T(\nu)$ by dividing it by $A$, the number of nucleons.  The 
nuclear Thomson term,\IE~the constant on the {\it r.h.s.} of Eq.~(\ref{eq:DR}) 
is given in terms of the fine structure constant $\alpha$, the net charge $Z$ 
of the target, and the mass of the nucleus given by $A$ times the nucleon 
mass, $M_N$ (in the following we ignore isospin breaking terms). The optical 
theorem relates the imaginary part of the Compton amplitude to the total 
photo-absorption cross section per nucleon $\sigma(\nu)$,
\bea
{\rm Im}T(\nu)&=&\frac{\nu}{4\pi}\sigma(\nu), \label{opt} 
\eea
so that the dispersion relation takes the form
\bea
{\rm Re}T(\nu)&=& - \frac{Z^2}{A^2} \frac{\alpha}{M_N}
 +\frac{\nu^2}{2\pi^2} \int_0^\infty\frac{d\nu'}{\nu'^2-\nu^2}\sigma(\nu'). 
\label{tt} 
\eea

To evaluate the dispersive integral, strictly speaking the photo-absorption 
cross section should be included all the way up to infinite energy; however, 
the scale separation between the nuclear and hadronic domains allows us to
approximate the integral by using a limited range of nuclear 
photo-absorption data.   
\begin{figure}
\includegraphics[height=6cm]{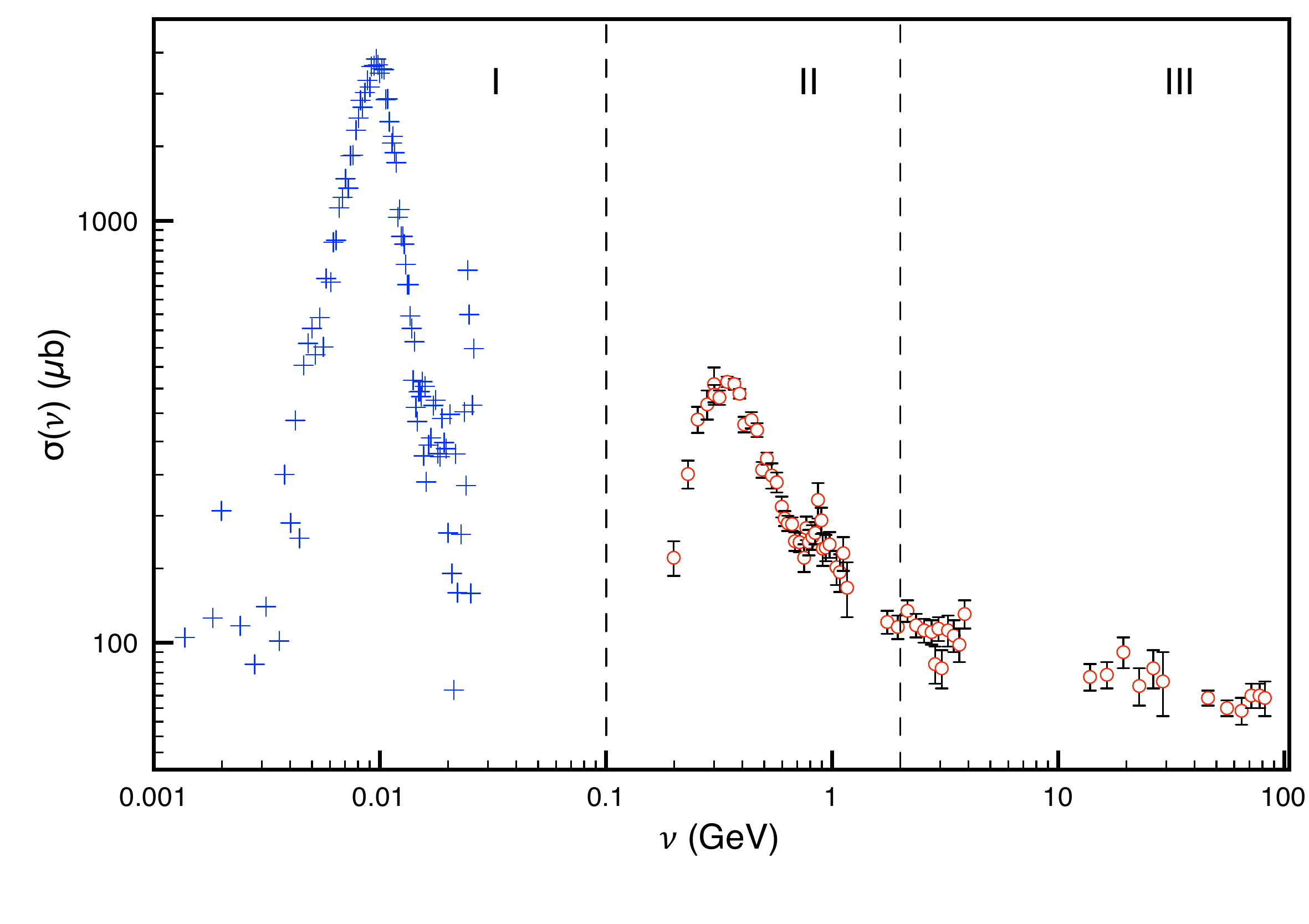}
\caption{
(Color online) Photo-absorption cross section data for a $^{207}$Pb target. 
Data in the nuclear range $\nu\leq27$ MeV (crosses) are from
\cite{Harvey:1964}; data in the hadronic and high energy range 0.2
GeV$\leq\nu\leq$100 GeV are from 
\cite{Hesse:1970cy,Caldwell:1973,Caldwell:1979,Bianchi}. 
Nuclear deformations are responsible for the giant resonance that saturates 
the cross section for $\nu \lesssim100$ MeV (region I). Excitations of 
individual nucleons are responsible for the hadronic resonances (region II) in 
the energy range between pion production threshold and 
${\cal O}(2-3 \mbox{ GeV})$. Finally for energies above a few GeV (region III),
the smooth cross-section is the result of partonic scattering via Regge 
exchanges.} 
\label{fig:regions}
\end{figure}
As shown in Fig.~\ref{fig:regions}, for a typical target nuclear resonances 
saturate the photo-absorption cross section for energies below 
$E_{\max} \approx 30\mbox{ MeV}$. The dominant feature of nuclear 
photo-absorption in the MeV range is the giant dipole resonance (GDR)
(cf. Ref. \cite{BermanFultz1975} for a comprehensive review of GDR data and 
theory).  As an example, the $^{207}$Pb data in the nuclear range 
are plotted along with the higher energy data in Fig.~\ref{fig:regions}, in 
which the GDR is seen as a sharp peak with width $\Gamma_{GDR} \approx 7$ MeV.
We evaluate the dispersion relation at $\nu_{max}  \lesssim  100$ MeV, which 
roughly demarcates the scale of hadronic physics where single-nucleon 
resonances begin contributing to the cross section,  
 \bea
{\rm Re}T(\nu_{max})&\approx& -\frac{Z^2}{A^2} \frac{\alpha
  }{M_N}-\frac{1}{2\pi^2}\int_{0}^{E_{max}}d\nu'\sigma(\nu'). \label{s1}
\eea
For an energy that is low compared to the hadronic scale, the scattering 
amplitude can be approximated by the sum of contributions describing photon 
interactions with point-like nucleons, \IE~it is given by a sum of Thomson 
terms on $Z$ protons,  
\bea
{\rm Re}T(\nu_{max})&\approx&-\frac{Z}{A}\frac{\alpha}{M_N} \ .
\label{eq:TRK_rhs}
\eea
Combining Eqs.~(\ref{s1}) and (\ref{eq:TRK_rhs}) leads to the
Thomas-Reiche-Kuhn  (TRK) sum rule~\cite{Levinger:1960}, 
(with $\alpha/M_N \approx 3.03\mbox{ mb~MeV}$), 
\begin{eqnarray}
\label{eq:TRK}
&& \int_{0}^{E_{max}} d\nu \sigma (\nu) = 2 \pi^2 \frac{NZ}{A^2} 
\frac{\alpha}{M_N} 
 \approx 60 \frac{NZ}{A^2} \mathrm{ mb~MeV}. \nonumber \\
\end{eqnarray}
Furthermore, adopting a Breit-Wigner form for the GDR cross section 
\begin{equation}
\label{eq:GDR}
\sigma(\nu)  \approx  \sigma_{\GDR} (\nu) = \frac{M_{\GDR}^2 \Gamma_{\GDR}^2\sigma_{\GDR}}
{(\nu^2 - M_{\GDR}^2)^2 + M_{\GDR}^2 \Gamma_{\GDR}^2},
\end{equation}
the integral over the resonance photo-absorption cross section gives  
$\pi\sigma_{\GDR}\Gamma_{\GDR}/2$, and the TRK sum rule leads to the relation
\bea
\sigma_{\GDR}\Gamma_{\GDR}  \approx 12\pi \frac{NZ}{A^2} \mbox{ mb~MeV}.
\label{eq:GDRb} 
\eea
In Eq.~(\ref{eq:GDRb}), $\sigma_{\GDR}$ is the value of the photo-absorption 
cross section at the peak of the GDR resonance, and $\Gamma_{\GDR}$ is the 
resonance half-width. This sum rule has been confronted with experimental data 
on a vast number of nuclear targets and is found to be satisfied to within 
$\sim30$\%.  This level of agreement demonstrates that the physics of nuclear
excitations is correctly described by a model assuming quasi-free nucleons
within the nucleus, which leads to Eq.~(\ref{eq:TRK_rhs}). However,
several model assumptions were used to equate the (model-dependent) 
r.h.s. of Eq.~(\ref{eq:DR}) to the integral over the total photo-absorption 
cross section. First, one assumes that the integral in the l.h.s. of
Eq.~(\ref{eq:TRK}) converges if the integration is extended to infinite 
energy, and in any case is dominated by the nuclear spectrum at 
$\nu\lesssim E_{max}$. Second, to eliminate the $\nu$-dependence one assumes 
that in  the expansion 
\begin{equation}
\nu^2\int_0^{E_{max}}\!\!\!\!\!\!d\nu'\frac{\sigma(\nu')}{\nu^2-\nu'^2}
= \left[1 + \frac{ \langle \nu^2 \rangle}{\nu^2} +\dots\right] 
\int_{0}^{E_{max}}\!\!\!\!\!\!d\nu'\sigma(\nu'),  
 \end{equation} 
the second term in the bracket, proportional to the mean squared energy 
averaged over the nuclear spectrum, satisfies 
\begin{equation} 
\langle \nu^2 \rangle=\frac{\int_{0}^{E_{max} }d\nu'
    \nu'^2\sigma(\nu')}{\int_{0}^{E_{max} }d\nu'\sigma(\nu')} \ll 1 \ . 
    \end{equation} 
 For typical values, $M_{\GDR}\sim15$ MeV,
$\Gamma\sim7$ MeV, $E_{max} = 30$ MeV and $\nu=\nu_{max}=100$ MeV one finds 
that $\langle \nu^2 \rangle /\nu_{max}^2$ amounts to a 10-15\% difference 
between the dispersive integrals in Eq.~(\ref{tt}) and Eq.~(\ref{s1}).   
With increasing  $\nu_{max}$ the correction term between the two integrals 
becomes smaller; however, the proximity of the pion production threshold and 
the nucleon excitation spectrum induces an important systematic error that 
cannot be accounted for within the framework of the TRK sum rule. These issues 
are addressed in the following section. 

\section{Nuclear photo-absorption in the range of nucleon resonances: a 
`constituent quark model' sum rule} 
\label{sec:j_form}
We now extend the arguments that lead to the TRK sum rule to energies in  
the nucleon excitation region, which we will define as the energy range 
between the threshold for pion production on a free nucleon, 
\IE~$\sim100 \mbox{ MeV}$ and a few GeV. For energies above $\nu_{max} = 
2-3\mbox{ GeV}$ the cross-section is smooth and does not exhibit resonance 
behavior. Above the resonance range we expect the cross section to be 
described by  scattering on individual  constituents of the nucleons 
\IE~constituent quarks.  
In analogy to Eq.~(\ref{eq:TRK_rhs}) we thus assume 
  \bea
{\rm Re}T_{\CQM} \equiv \mbox{Re}T(\nu_{max})  &\approx&- \frac{1}{A} 
\sum_{q\in A} \frac{\alpha}{m_q}   e_q^2 \nonumber \\ & = &  
- \frac{3Z + 2N}{A} \frac{\alpha}{M_N}. 
\label{eq:TRK_rhs1}
\eea
Following the derivation of the TRK sum rule we want to identify 
$\mbox{ReT}_{\CQM}$ with the sum of the nuclear Thomson term and the 
photo-absorption cross-section integrated up to some energy above the nucleon 
resonance region. This is complicated by the fact that above the resonance 
region, the photo-absorption cross section does not fall off with energy but 
instead increases until it is close to the Froissart bound \cite{Froissart:1961}. 
 This increase with energy occurs because in QCD the photon does not interact 
with a fixed number of hadron constituents (\EG~the nucleon, pion, constituent 
quarks) but as the beam energy increases gluon showers build up between the 
photon and the target. Phenomenologically one describes this energy region 
in terms of Pomeron exchange. Furthermore, at intermediate energies before 
the universal Pomeron scattering takes over, Reggeon exchanges \IE~parton 
showers dominated by exchange of quarks, contribute a significant background 
to hadron resonance production. Since it is only the hadron resonances  
that can be associated with constituent quark degrees of freedom, a sum rule 
involving constituent quarks must involve cross sections with both the Reggeon 
and the Pomeron contributions subtracted. 
The Regge and Pomeron contributions to the cross section per nucleon are 
conventionally parametrized by 
\begin{equation} 
\sigma^{R+P}(\nu)  = \sigma^R_T  + \sigma^P_T = \sum_{i=R,P} c_i \left( \frac{\nu}
{\mbox{ GeV}} \right)^{\alpha_i(0)-1} \ ,  
\label{regge} 
\end{equation} 
where for the Regge and Pomeron contributions we use the intercepts 
$\alpha_R(0) = 1/2$ and $\alpha_P(0) = 1.097$, respectively 
\cite{Breitweg:1999}. This corresponds to an amplitude given by 
  \begin{eqnarray}
\label{eq:Regge}
T^{R+P}(\nu)  = T^R + T^P   &= &  - \sum_{i=R,P} \frac{c_i}{4\pi}\frac{1 + e^{-i \pi \alpha_i (0)}}
  {\sin \pi \alpha_i (0)} \nu^{\alpha_i (0)} \nonumber \\
  & = & \frac{\nu^2}{2\pi^2} \int_0^\infty \frac{d\nu'}{\nu'^2 - \nu^2} \sigma^{R+P}(\nu') 
\nonumber \\  \label{R+P} 
\end{eqnarray}
Eq.~(\ref{eq:DR}) can now be rewritten by adding and subtracting  the 
asymptotic contributions given by Eqs.~(\ref{regge}) and (\ref{R+P}) to 
obtain  
\begin{eqnarray}  
  \mbox{Re} T(\nu) & = &  - \frac{Z^2}{A^2} \frac{\alpha}{M_N}  \nonumber \\
    & + & \frac{\nu^2}{2\pi^2}  \int_0^\infty d\nu' 
  \frac{ \sigma(\nu')  - \sigma^{R+P}(\nu') }{\nu'^2 - \nu^2}  
  +  \mbox{Re} T^{R+P} (\nu) \ . 
\nonumber \\ 
\label{DR-sub} 
\end{eqnarray} 
In Eq.~(\ref{DR-sub}) the integrand on the {\it r.h.s} vanishes asymptotically 
and we can take the limit $\nu \to \infty$ to obtain 
\begin{eqnarray}
& & 
\lim_{\nu \to \infty} \frac{\nu^2}{2\pi^2}  \int_0^\infty d\nu' 
\frac{ \sigma(\nu')  - \sigma^{R+P}(\nu') }{\nu'^2 - \nu^2}  
 =   \nonumber \\
& =  &  -  \frac{1}{2\pi^2}  \int_0^{E}  d\nu'   \sigma(\nu')  
+   \sum_{i=R,P} \frac{c_i  \mbox{ GeV} }{2\pi^2\alpha_i(0) } 
\left( \frac{E}{\mbox{ GeV}} \right)^{\alpha_i(0)} 
\nonumber \\ \label{limit} 
\end{eqnarray} 
In Eq.~(\ref{limit}) $E$ is the energy above which we can neglect the 
difference between the data $\sigma(\nu)$ and the high-energy asymptotic form 
$\sigma^{R+P}(\nu)$. We used $E=2$GeV in our calculations.
To extend the TRK sum rule to energies above the nucleon 
resonance region we postulate that the contribution to the {\it r.h.s} of 
Eq.~(\ref{DR-sub}) from the photo-absorption cross section in the nucleon 
resonance region, reduced by the Regge plus Pomeron background can be 
represented by Thomson scattering on the constituent quarks. This leads to a 
phenomenological finite-energy sum rule 
\begin{eqnarray}
\mbox{Re}T_{\CQM}   =     - \frac{Z^2}{A^2} \frac{\alpha}{M_N} 
 &  - &  \frac{1}{2\pi^2}  \int_0^{E}  d\nu' \sigma(\nu') \nonumber \\ 
 &  +  & \frac{c_R \mbox{ GeV}}{2\pi^2\alpha_R(0) } 
 \left( \frac{E}{\mbox{ GeV}} \right)^{\alpha_R(0)}
\nonumber \\ 
\end{eqnarray} 
or 
\begin{eqnarray} 
  - \left( 2 +\frac{ZN}{A^2} \right)\frac{\alpha}{M_N}    & =   &  
   -  \frac{1}{2\pi^2}  \int_0^{E}  d\nu'   \sigma(\nu')  \nonumber \\
 &   + &  \frac{c_R  \mbox{ GeV}}{2\pi^2\alpha_R(0) } \left( \frac{E}{\mbox{ GeV}} \right)^{\alpha_R(0)}  
\nonumber \\
\label{cqm} 
\end{eqnarray}

In addition to the phenomenological $CQM$ sum rule, the dispersion relation 
in  Eq.~(\ref{DR-sub}) can also be used to calculate the value of the 
sub-leading energy-independent contribution to the photo-nuclear Compton 
amplitude, \IE~the $\alpha = 0$ pole discussed in Sec.~\ref{sec:intro}.  
\begin{eqnarray}  
& &  \mbox{Re} T^{\alpha=0} \equiv \lim_{\nu \to \infty}  [\mbox{Re} T(\nu) - \mbox{Re} T^{R+P}(\nu)]  =  
    - \frac{Z^2}{A^2} \frac{\alpha}{M_N}   \nonumber \\ 
 & &  -  \frac{1}{2\pi^2}  \int_0^{E}  d\nu'   \sigma(\nu')  
   + \sum_{i=R,P} \frac{c_i  \mbox{ GeV} }  {2\pi^2\alpha_i(0) } 
  \left( \frac{E}{\mbox{ GeV}} \right)^{\alpha_i(0)} 
\nonumber \\ 
\label{j0} 
\end{eqnarray} 
From Eq.~(\ref{j0}) we see that the $\alpha=0$ pole contribution is given 
by the difference between the full scattering amplitude and the 
contribution to the scattering amplitude from Regge plus Pomeron terms. 

\subsection{Numerical results} 
To compute the integral over the photo-absorption cross section we parametrize the hadronic 
cross-section by a sum of up to 6 Breit-Wigner resonances plus a smooth background, 
%
\begin{equation}
\sigma(\nu) =
\sum_{i=1}^6\frac{M_i^2}{s}\frac{\sigma_iM_i^2\Gamma_{tot,i}\Gamma_{\gamma,i}}
  {(s-M_i^2)^2+M_i^2\Gamma_{tot,i}^2} + \sigma^{R+P}(\nu) \ . 
\label{eq:Res}
\end{equation}
In Eq.~(\ref{eq:Res}), $s=M^2+2M\nu$ is the square of the c.m. energy. 
In order of increasing mass, we account for the following resonances:   
$P_{33}(1232)$, $P_{11}(1440)$, $D_{13}(1520)$, $S_{11}(1665)$,
$F_{15}(1680)$, and $F_{37}(1950)$. We use energy-dependent widths,
\beqn
\Gamma_{\gamma,i}&=&\Gamma_i\left[\frac{1+X^2/K_i^2}{1+X^2/K^2}\right]^{J_\gamma}\nn\\
\Gamma_{tot,i}&=&\Gamma_i\frac{q}{q_i}\left[\frac{1+X^2/q_i^2}{1+X^2/q^2}\right]^{l},
\eeqn
where $K$ and $q$ are respectively the momenta of the photon and single pion decay
in the c.m. frame, and $K_i$ and $q_i$ refer to their values at
the resonance position $\sqrt{s}=M_i$. $J_\gamma$ is the spin of the resonance, $l$ is the 
angular momentum of the decay products, and $\Gamma_i$ is the intrinsic width of 
the $i$th resonance. 
The damping parameter $X$ was set to $X=0.15$ GeV for the $P_{33}(1232)$
and we chose $X=0.35$ GeV for all other resonances.

\begin{figure*}
\includegraphics[height=13cm]{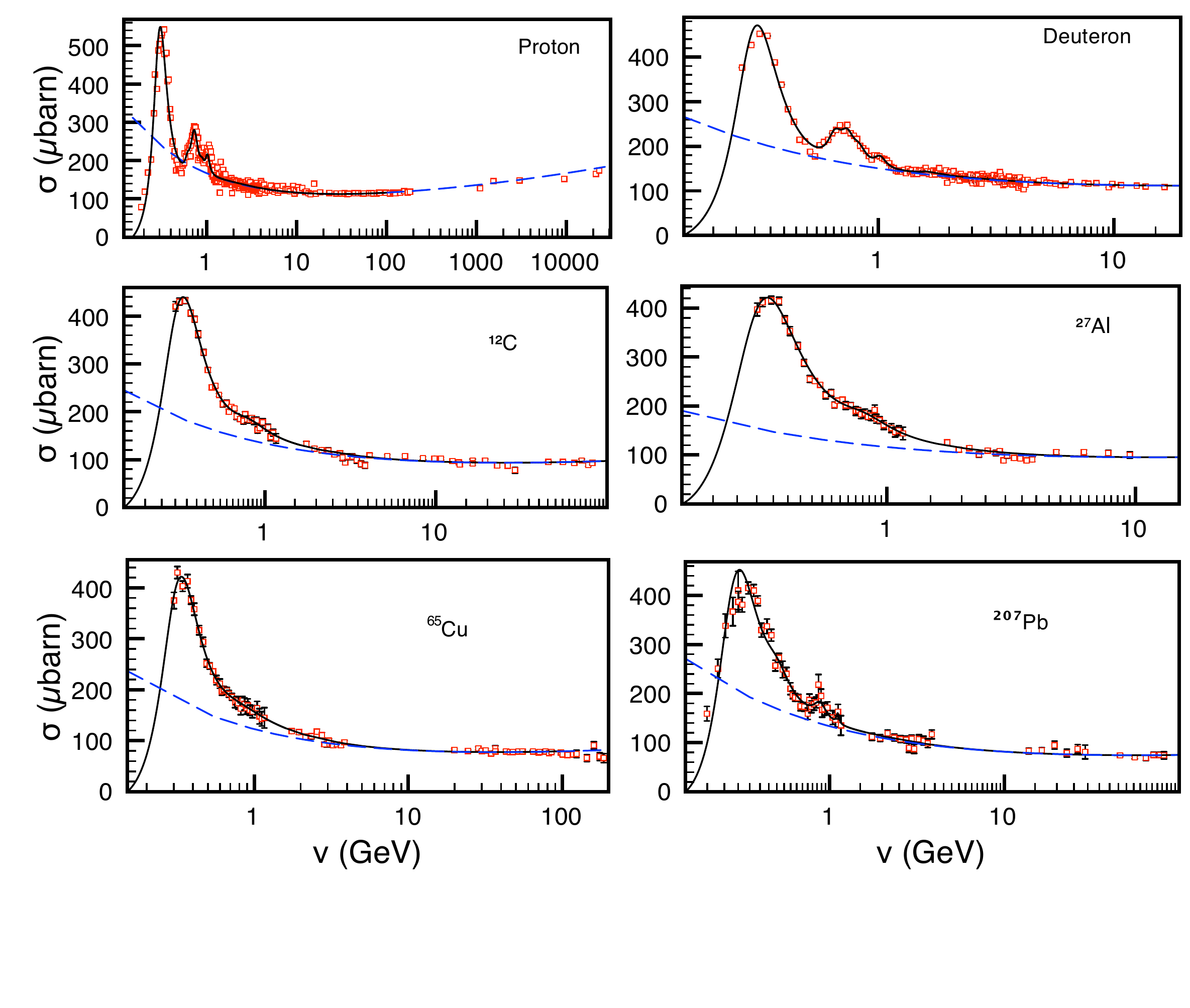}
\vspace{-2cm}
\caption{(Color online) High energy photo-absorption cross sections per nucleon 
for six nuclear targets compared to the fit results (solid lines) using the Breit-Wigner 
resonance plus background pametrization of Eq.~(\ref{eq:Res}). Data are from \cite{PDG} for 
the proton and the deuteron, and from \cite{Caldwell:1973,Caldwell:1979,Bianchi} for heavier 
nuclei. The Regge plus Pomeron curves are shown by dashed lines. The background 
fit parameters are 
given in Table \ref{tab_reggefit}. }
\label{fig:HE}
\end{figure*}

The Regge plus Pomeron background is chosen so that it explicitly matches onto the Regge
cross section, 
\beqn
\sigma^{R+P}(\nu)&=&f_{thr}\left[A
  \left(\frac{\nu}{1\,GeV}\right)^{0.097}+B \left(\frac{\nu}{1\,GeV}\right)^{-0.5}\right]\nn\\ 
\label{eq:backg}
\eeqn
where the threshold factor $f_{thr}$ in Eq.~(\ref{eq:backg}) ensures that the
background cross section vanishes at pion production threshold, 
\beqn
f_{thr}&=&1-e^{-\frac{2(\nu-\nu_\pi)}{M}},
\eeqn
as proposed in \cite{Bianchi}. 
Rather than using the Regge intercepts as free parameters, we take those intercepts from 
fits to photo-absorption data on the proton \cite{Breitweg:1999}. This is partly motivated by 
the limited energy range over which nuclear data are available. 
\begin{table*}
\vspace{1cm}
   \begin{tabular}{|l|l|l|l|l|l|l|}
\hline
&Proton & Deuteron & $^{12}_6$C & $^{27}_{13}$Al & $^{65}_{29}$Cu & $^{207}_{82}$Pb \\
\hline
$c_P$ ($\mu$b) & $68.0\pm0.2$ & $70.08\pm1.26$ & $57.24\pm1.13$ & $62.70\pm6.0$ & $45.88\pm0.57$ & $42.08\pm1.96$ \\
\hline
$c_R $ ($\mu$b) & $99.0\pm1.15$ & $80.50\pm2.27$ & $76.49\pm4.40$ & $53.53\pm11.6$ & $76.95\pm3.60$ & $91.43\pm9.14$ \\
\hline
  \end{tabular}
\caption{Reggeon and Pomeron parameters in $\mu b$}
\label{tab_reggefit}
\end{table*}
 In Fig. \ref{fig:HE} we display the results of the cross section fits using 
Eq.~(\ref{eq:Res}) (solid lines) along with the Regge + Pomeron background (dashed lines). 
The parameters of the background cross sections are listed in Table \ref{tab_reggefit} for 
each nuclear target. 

We note that in general due to nuclear effects such as resonance
broadening and Fermi motion, the division of the cross section into
resonance plus background becomes somewhat ambiguous. Examining 
Fig. \ref{fig:HE} we notice that for heavier nuclei, as
opposed to the proton and to some degree the deuteron, there is no clear 
resonance structure around and above 
$\nu=1$ GeV. For heavier nuclei the effect of broadening and overlapping of 
resonances can alternatively be reproduced by enhancing the Reggeon strength $c_R$, 
thus the strength of the Regge background is sensitively correlated with the 
choice of the resonance parameters. 

Uncertainties in identification of the Pomeron and Reggeon background for nuclear targets 
produces error bars for those parameters that are significantly larger than those 
parameters for the proton. In particular, for $^{27}$Al the highest energy data available are 
as low as 10 GeV and the corresponding fit does not allow for a precise determination of the 
background parameters, and this is reflected by the large errors in Table \ref{tab_reggefit}.
Our fits provide a reduced $\chi^2$ per degree of freedom of order one, except for the 
Aluminum target, where it was greater than two.
 
\begin{figure*}
\includegraphics[width=9cm]{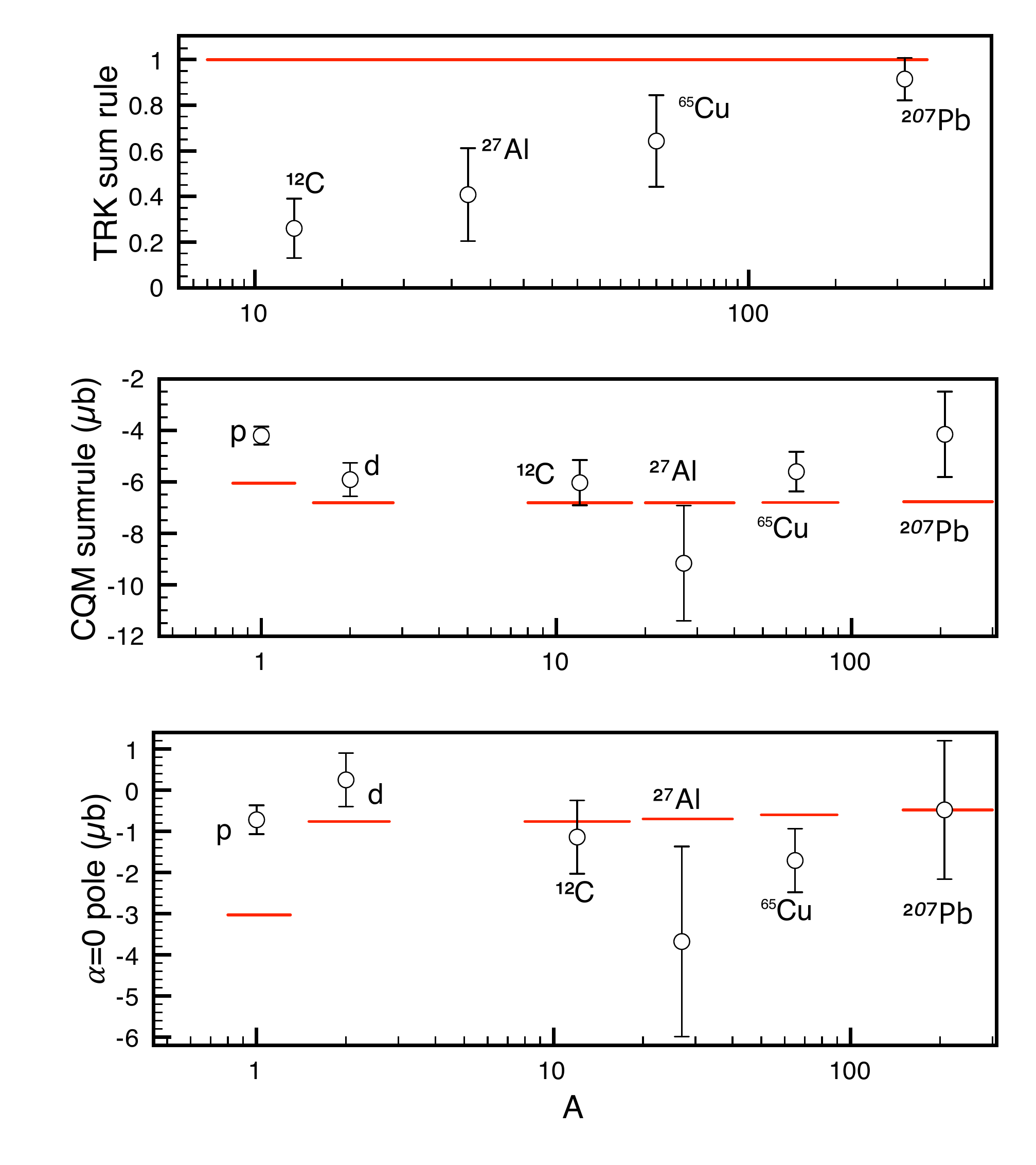}
\caption{(Color online) Upper panel: the fraction of the TRK sum rule for nuclear targets $^{12}$C, $^{27}$Al,
  $^{65}$Cu, and $^{207}$Pb; middle panel: experimental values (data points) vs.~theoretical 
expectation (dotted line) for our new constituent quark model (CQM) sum rule for the proton, 
deuteron, $^{12}$C, $^{27}$Al, $^{65}$Cu, and $^{207}$Pb, in units of $\mu$b; lower panel: results 
for the $\alpha=0$ pole for all targets considered, in $\mu$b.}
\label{fig:Fig4}
\end{figure*}

In Table \ref{tab} we list the numerical values of the various contributions to the 
constituent quark sum rule in Eq.~(\ref{cqm}), and the $\alpha=0$ pole contribution given 
in Eq.~(\ref{j0}). A comparison of theoretical and experimental contributions to the finite 
energy sum rules and the $\alpha=0$ pole contribution are displayed in Fig. \ref{fig:Fig4}. 
The upper panel shows the comparison of data in the nuclear energy range with the TRK sum 
rule for four nuclear targets (we plot the fraction of the TRK sum rule for each nucleus); 
the middle panel shows the comparison of data in the nuclear 
and hadronic energy region with the predictions from our new constituent quark model or 
CQM sum rule (this is given by the fourth and fifth rows of Table \ref{tab}); finally, the 
lower panel shows the predictions of dispersion relations for the 
value of the $\alpha=0$ pole for the six nuclear targets in comparison
with the corresponding Thomson term values (the final and second-last row in Table 
\ref{tab}, respectively). 

We observe that the CQM sum rule is better obeyed for heavier nuclei than for the proton or 
deuteron. One possible explanation could be that this sum rule amounts to counting the 
effective number of quarks within the target; this relies on a mean-field approach to the 
target which we would expect to become more accurate as the number of target nucleons 
increases. For the $\alpha=0$ pole contribution, our new result for
the proton is significantly different from the Thomson term, 
which is at variance with the original result of Damashek and 
Gilman~\cite{Damashek:1969xj}. This discrepancy is due to our use of the very high energy 
photo-absorption data that has become available only recently \cite{Breitweg:1999}.
As a result, instead of the high-energy parameterization used in Ref.~\cite{Damashek:1969xj}, 
 
\beqn
\sigma^{R+P}(\nu)\approx\left(96.6+70.2\sqrt{\frac{1\,{\rm
        GeV}}{\nu}}\right)\mu {\rm b},
\label{eq:RPold}
\eeqn
we find 
\beqn
\sigma^{R+P}(\nu)\approx\left(68.0
\left[\frac{\nu}{1\,{\rm GeV}}\right]^{0.097}\!\!\!+99.0\sqrt{\frac{1\,{\rm
        GeV}}{\nu}}\right)\mu {\rm b}.
\label{eq:RPnew}
\eeqn
At an energy $\nu=1$ GeV both formulas give almost identical results, but at high energies 
they differ dramatically. At the same time, the data in the resonance region have not changed 
much, so this leads to our new value for the $\alpha=0$ contribution to 
photo-absorption on the proton. 

For heavier nuclei, however, the bottom panel of Fig.~\ref{fig:Fig4} and the final row  
of Table \ref{tab} shows that the $\alpha=0$ contribution appears to
be consistent with the Thomson term. This result 
is due to an interplay of various nuclear effects in the resonance region that affect the value of 
the integrated photo-absorption cross section, and also shadowing at medium-to-high energies. 
Shadowing at energies below $\nu=200$ GeV causes the value of $c_P$ to decrease from 68 
$\mu$b for the proton to approximately 43 $\mu$b for lead, respectively. On the other hand, 
the Pomeron is a QCD phenomenon that is due to the interaction of quarks and gluons and 
should be the leading mechanism of photoabsorption at extremely high energies. It can 
be expected that at asymptotic energies nuclear effects should be
negligible, and the strength of the Pomeron should be the same for both the
proton and heavier nuclei.  If in the future nuclear photoabsorption 
data above $\nu=200$ GeV would become available, they could shed more light on the
asymptotic behavior of the forward nuclear Compton amplitude, and could remove 
uncertainties regarding the strength of the Pomeron, Reggeon and $\alpha=0$ pole 
contributions.

\color{black}
\begin{table*}
\vspace{1cm}
   \begin{tabular}{|l|l|l|l|l|l|l|}
\hline
&Proton & Deuteron & $^{12}_6$C & $^{27}_{13}$Al & $^{65}_{29}$Cu & $^{207}_{82}$Pb \\
\hline
$\frac{1}{2\pi^2A}\sigma_{int}^{had}$ & $18.60\pm0.31$ & $17.46\pm0.51$ & $16.80\pm0.62$ & $16.54\pm1.50$ & $16.16\pm0.57$ & $16.57\pm1.02$ \\
\hline
$\frac{1}{2\pi^2A}\sigma_{int}^{nucl}$& - & - & 0.197 & 0.30 & 0.480 & 0.69 \\
\hline
$\frac{1}{2\pi^2} c_{R} \frac{(E/GeV)^{1/2}}{1/2}$& $14.19\pm0.16$& $11.54\pm0.39$ & $10.96\pm0.63$  & $7.67\pm1.66$  &  $11.03\pm0.52$  & $13.10\pm1.31$\\
\hline
 r.h.s of Eq. (17) & $-4.21\pm0.35$ & $-5.92\pm0.65$ & $-6.04\pm0.88$  & $-9.17\pm2.24$  &  $-5.61\pm0.77$  & $-4.16\pm1.66$\\
\hline
$-\left(2+\frac{ZN}{A^2}\right)\frac{\alpha}{M}$& $-6.06$ & $-6.82$ & $-6.82$ & $-6.82$ & $-6.81$ & $-6.78$ \\
\hline
$\frac{1}{2\pi^2} c_{P} (E/GeV) $& $6.72\pm0.02$ & $6.92\pm0.12$ & $5.65\pm0.11$ & $6.19\pm0.59$ & $4.53\pm0.06$ & $4.16\pm0.25$ \\
\hline
$-\frac{Z^2}{A^2}\frac{\alpha}{M}$& $-3.03$ & $-0.76$ & $-0.76$ & $-0.70$ & $-0.60$ & $-0.48$ \\
\hline
 Re$T^{\alpha=0}$ & $-0.72\pm0.35$ & $0.25\pm0.65$ & $-1.14\pm0.89$  & $-3.68\pm2.31$  &  $-1.71\pm0.77$  & $-0.48\pm1.68$\\
\hline
   \end{tabular}
\caption{
  Contributions to the finite energy sum rule for selected targets in units of 
GeV$\cdot\mu$b. The entries in the second row are taken from a
review on nuclear data in Ref. \cite{BermanFultz1975}.
}
\label{tab}
\end{table*}

Finally, in addition to the paper by Damashek and Gilman Ref. \cite{Damashek:1969xj} already 
mentioned above, there have been other evaluations of the $\alpha=0$ pole for forward Compton 
scattering. Dominguez, Ferro Fontan and Suaya \cite{Dominguez:1970wu} 
and Shibasaki, Minamikawa and Watanabe \cite{Shibasaki:1971} used 
a similar approach to that of Ref. \cite{Damashek:1969xj} and
independently arrived at a qualitatively similar result, 
\beqn
{\rm Re}T_p^{\alpha=0}=(-3\pm2) \mu{\rm b\,GeV},
\eeqn
where the uncertainty is dominated by the parameters of the
high energy fit, reflecting the limited range of high-energy data available at that
time.

In Ref. \cite{Dominguez:1972}, Dominguez, Gunion and Suaya extended
this analysis by including the deuteron photoabsorption data. They
employed a model for nuclear effects to extract parameters of the
neutron from deuteron and proton data, and evaluated the FESR for both
nucleons. Their conclusions were that the $\alpha=0$ pole is
consistent with the respective Thomson term for both,
\beqn
{\rm Re}T_n^{\alpha=0}&=&(0\pm1.5) \mu{\rm b\,GeV},\nn\\
{\rm Re}T_p^{\alpha=0}&=&(-3\pm0.8) \mu{\rm b\,GeV},\label{eq:j=0dominguez}
\eeqn
where ${\rm Re}T_{p(n)}^{\alpha=0}$ refers to the proton (neutron),
respectively. Tait and White in Ref. \cite{Tait:1972} re-analyzed the FESR
using a more recent data set, and obtained a much more conservative
estimate, 
\beqn
{\rm Re}T_p^{\alpha=0}=(-3^{+4}_{-5})\mu{\rm b\,GeV}.
\eeqn
Based on the recent proton data on photoabsorption at
very high energies \cite{Breitweg:1999} and the analysis of Tait and White \cite{Tait:1972}, 
we conclude that the errors in Eq. (\ref{eq:j=0dominguez}) were
significantly underestimated.

\section{Summary and Conclusions}
\label{summary}  
In summary, we revisited the finite energy sum rules (FESR) for forward
real Compton scattering on the proton and heavier nuclei. As the photon energy 
increases and its wavelength decreases, the Compton amplitude becomes sensitive 
to progressively smaller features of a nuclear target. At the lowest 
energies, the Compton amplitude is determined by scattering on the
target as a whole, whereas in the high energy limit it is expected to
be determined by scattering on elementary target constituents. 

Finite energy sum rules provide a qualitative comparison between the high energy 
and the low energy limits of the scattering amplitude. For nuclei, the Thomas-Reiche-Kuhn (TRK) 
sum rule relates the strength of the giant dipole resonance to the
difference between the nuclear Thomson term and the incoherent sum of
Thomson terms of protons residing in the nucleus. In a similar fashion we have proposed a new 
sum rule that describes the integrated strength of the nucleon resonances as a difference between
the nuclear Thomson term and the incoherent sum of Thomson terms from constituent quarks residing 
in the target.  In this process it is crucial to separate the Reggeon and Pomeron high-energy 
contributions to the FESR, and we call this new sum rule the `constituent quark model' or CQM 
sum rule.

We analyzed the TRK and CQM sum rules for the proton, deuteron, $^{12}$C,
$^{27}$Al, $^{65}$Cu and $^{207}$Pb targets.  All nuclear data are
consistent with the CQM sum rule, however, for the proton the
comparison is not as favorable. This may be explained by the fact
that in a nucleus, the errors due to various systematic effects are  
averaged over a large number of constituent quarks, and thus may be 
statistically less significant than for the proton. 

Theoretical arguments suggest that Compton scattering amplitudes at high energies should contain an 
energy-independent constant that corresponds to a Regge pole at $\alpha=0$
~\cite{Damashek:1969xj,j0r,j01,Creutz:1968ds,Close:1971ed,Zee:1972nq, Brodsky:1971zh, Brodsky:1973hm,Brodsky:2008qu}.
Previous attempts to extract this constant obtained results consistent with this amplitude 
being approximately equal to the value at low energies \IE~the Thomson term. 
 
We were able to demonstrate that high-energy photoabsorption data on the proton confirms that the
$\alpha=0$ pole and the Thomson term, Re$T_p(0)=-3.03 \mu$bGeV are significantly different, 
\beqn
{\rm Re}T_p^{\alpha=0}&=&(-0.72\pm0.35) \mu{\rm b\,GeV}.
\eeqn

The difference between our result and the value 
consistent with the Thomson term from previous analyses 
\cite{Damashek:1969xj,Dominguez:1970wu,Shibasaki:1971,Tait:1972} is due to the recent high-energy 
photoabsorption data \cite{Breitweg:1999}, which changes the Regge plus Pomeron 
contribution from Eq.~(\ref{eq:RPold}) which was used in Ref.~\cite{Damashek:1969xj}, 
to our background form Eq.~(\ref{eq:RPnew}). With this form for the background amplitude 
we extract the new value for the $\alpha=0$ pole that differs
significantly from the value of the Thomson term.
 
We extended this analysis to a number of nuclear targets. 
In the case of the sum rule for the $\alpha=0$ pole, only the proton
result is unambiguously distinct from the Thomson term, whereas for other targets
the result was consistent with the Thomson term within the experimental errors. 
Our results are relevant to the question of the $A$-dependence of the Pomeron contribution. The 
Pomeron in QCD is isospin-independent, and for asymptotically high energies one 
generally expects that the Pomeron contribution from a free proton should equal the average
nucleon Pomeron contribution in a nucleus. Current nuclear data only extend up
to $\sim200$ GeV and at these energies shadowing effects are responsible
for suppressing the Pomeron contribution by some 30\% relative to the proton
value. It is an open question whether future nuclear photo-absorption data at higher energies 
similar to those currently available for the proton will tend to bring the Pomeron strength
per nucleon up to the free proton value.

\section*{Acknowledgments}

The authors would like to thank Stan Brodsky for numerous discussions. This 
work was supported in part by the US Department of Energy under 
contract DE-FG0287ER40365 and the US National Science Foundation under grant 
PHY-0854805. 


%

%

\begin{thebibliography}{99}
\bibitem{footnote} The difference between the notation $\alpha=0$ pole
  and $J=0$ pole, more commonly used in the literature, is a subtle one. $J$
  refers to the spin of an exchanged particle, whereas $\alpha$ refers
  to a parameter describing the high-energy behavior of a scattering
  amplitude, the intercept of a Regge trajectory. For mesonic trajectories,
  $\alpha=0,1,2,\dots$ corresponds to physical particles of integer spin
  $J$. Instead, the $J=0$ pole cannot be identified with such 
an exchange, therefore, we opt for the notation $\alpha=0$ pole 
as more suitable.
\end{thebibliography}
\end{document}